\def\year{2020}\relax
\documentclass[letterpaper]{article} 
\usepackage{aaai20}  
\usepackage{times}  
\usepackage{helvet} 
\usepackage{courier}  
\usepackage[hyphens]{url}  
\usepackage{graphicx} 
\usepackage{booktabs} 
\usepackage{multirow}
\usepackage{xcolor}
\usepackage{amsfonts}
\usepackage{amsmath}
\usepackage{algorithm2e}
\usepackage{subfigure}
\usepackage{graphicx}  
\title{Inter-sequence Enhanced Framework for Personalized Sequential Recommendation}
\author{Feng Liu\textsuperscript{\rm 1}, Weiwen Liu\textsuperscript{\rm 2}, Xutao Li\textsuperscript{\rm 1}, Yunming Ye\textsuperscript{\rm 1} \\
\textsuperscript{\rm 1}Harbin Institute of Technology\\ 
fengliu@stu.hit.edu.cn, \{lixutao, yeyunming\}@hit.edu.cn \\
\textsuperscript{\rm 2}The Chinese University of Hong Kong\\ 
wwliu@cse.cuhk.edu.hk
}

\begin{document}

\maketitle

\begin{abstract}
Modeling sequential correlation of users' historical interactions is essential in sequential recommendation.
However, the majority of the approaches mainly focus on modeling the \emph{intra-sequence} item correlation within each individual sequence but neglect the \emph{inter-sequence} item correlation across different user interaction sequences. 
Though several studies have been aware of this issue, their method is either simple or implicit.
To make better use of such information,
we propose an inter-sequence enhanced framework for the Sequential Recommendation (ISSR). In ISSR, both inter-sequence and intra-sequence item correlation are considered. Firstly, we equip graph neural networks in the inter-sequence correlation encoder to capture the high-order item correlation from the user-item bipartite graph and the item-item graph. Then, based on the inter-sequence correlation encoder, we build GRU network and attention network in the intra-sequence correlation encoder to model the item sequential correlation within each individual sequence and temporal dynamics for predicting users' preferences over candidate items.
Additionally, we conduct extensive experiments on three real-world datasets. The experimental results demonstrate the superiority of ISSR over many state-of-the-art methods and the effectiveness of the inter-sequence correlation encoder. 

\end{abstract}



\section{Introduction}


Sequential recommendation (SR) is a vital task in recommender systems. SR aims at predicting successive items that the user is likely to interact with by modeling sequential and transitional correlation~\cite{RendleFpmc,he2016fusing,HidasiKBT15Gru4rec,Tangcaser,wangchengkangSelfattention,MaHGN}. 






Modeling the item correlation within a user's sequence of interacted items or across different users' sequences of interactions lies at the core of modern SR~\cite{HidasiKBT15Gru4rec,Tangcaser,wangchengkangSelfattention,MaHGN}, namely intra-sequence and inter-sequence item correlation. 
Figure 1 shows an intuitive example from MovieLens, concretely, we term the \textbf{intra-sequence item correlation} as the sequential dependence of two items within a user's sequence of interacted items. 
For example, movie 1721 and 1682 are intra-correlated (with order 1) because they are sequentially correlated in User 8's behavior sequence.
We term \textbf{inter-sequence item correlation} as the occurrence of two items in difference users' sequences and there exists a path between these two items with some intermediate nodes.
For example, movie 1721 and 1784 are inter-correlated (with order 4) because they exist in user8's and user 201's sequences and there exist a path (${v_{1721}} \to {u_8} \to {v_{1357}} \to {u_{201}} \to {v_{1784}}$) between them.
In fact, we observe in the log data \footnote{https://grouplens.org/datasets/movielens/1m/} that user 90 actually watches movie 1682 and 1784 after she watched movie 1721.
This observation verifies that both intra- and inter-sequence item correlation are informative for SR.

\begin{figure}[!t]
\center
\includegraphics[width=0.45\textwidth]{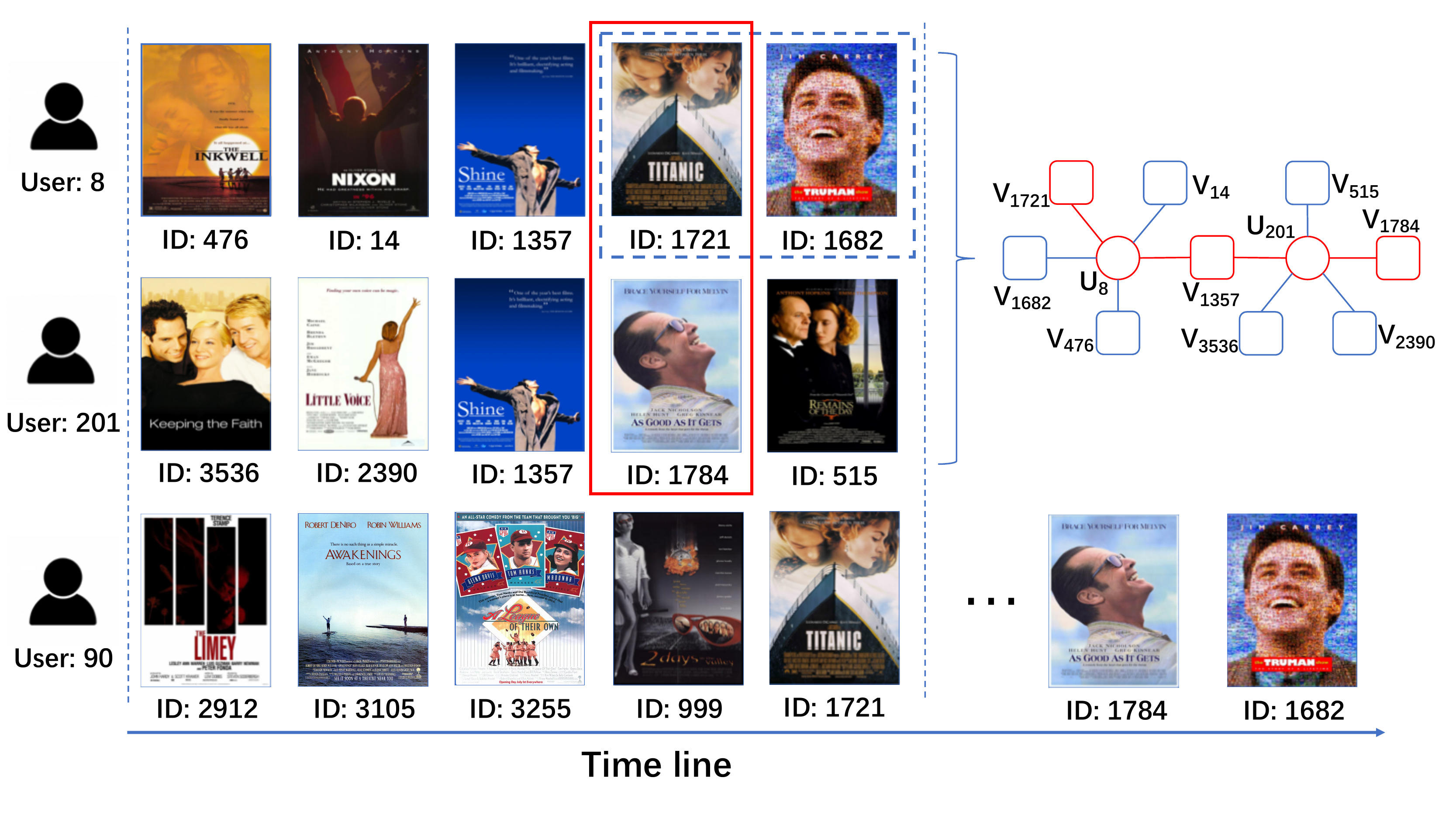}
   \vspace{-3ex}
\caption{\small{Inter- and intra-sequence behaviors in MovieLens. (The movie 1721 and 1682 are intra-sequence correlated in blue dashed box, and the movie 1721 and 1784 are inter-sequence correlated in red box.)}}
\label{fig:sequence_behavior}
\end{figure}

However, existing works for SR mainly put effort in modeling intra-sequence correlation yet neglect the effect of inter-sequence correlation. The authors in~\cite{HidasiKBT15Gru4rec,DBLP:conf/recsys/QuadranaKHC17,hidasi2018recurrent} apply recurrent neural networks (RNNs), which aggregate the sequence of the user's interacted items to capture sequential correlation among the items.
Different from RNN-based approaches, \cite{Tangcaser,yuanfajiecnn,xu2019recurrent} treat the embedding matrix of items in a sequence as an image and apply convolutional neural networks (CNNs) to model sequential correlation. However, the above RNN-based and CNN-based approaches do not model the different impacts of the items consumed at different time steps on current decision. 
Hence, the authors in \cite{wangchengkangSelfattention,sun2019bert4rec,zhang2019next} adopt attention networks to differentiate and learn the contribution of each individual item in a sequence to model the user interest when making predictions on next items. 
To obtain accurate item embedding and take complex transitions of items into account, Wu et al.~\cite{shuwuSrgnn} propose SR-GNN for session-based recommendation.
In addition, gated networks~\cite{MaHGN} and neural variational models~\cite{DBLP:conf/www/XiaoLM19} are also utilized in SR.
However, the above mentioned methods mainly focus on modeling the intra-sequence item correlation within each individual sequence and the inter-sequence item correlation across different sequences is neglected. 
Though the intra-sequence item correlation is vital, we argue that explicitly modeling inter-sequence item correlation is also critical, as it not only captures the users' general tastes, but also can help remedy the data scarce issue in SR\cite{he2016fusing}.



To the best of our knowledge, only a few existing studies improve the recommendation quality by considering both intra- and inter-sequence item correlation in SR~\cite{RendleFpmc,he2016fusing,DBLP:conf/sigir/WangRMCMR19}. FPMC \cite{RendleFpmc} applies a first-order Markov Chain (MC) to model users’ sequential behavior and utilizes matrix factorization to learn the inter-sequence item correlation. 
Later, Fossil is proposed to address the data scarce problem by utilizing high-order Markov chain and similarity models~\cite{he2016fusing}.
Recently, Wang et al. propose Collaborative Session-based Recommendation Machine (CSRM) and consider inter-sequence correlation between different sessions.  
However, the inter-sequence information is not fully exploited in these works: 
(i) For FPMC and Fossil, the order of the inter-sequence item correlation is limited by the latent method, which is the MF and similarity model respectively;
(ii) For CSRM, the inter-sequence correlation is considered in an implicit way, a simple session-level nearest neighbor-based approach.

Therefore, to make better use of inter-sequence item correlation, in this paper, we propose the \textbf{I}nter-\textbf{S}equence enhanced framework for personalized \textbf{S}equential \textbf{R}ecommendation (ISSR), where both intra and inter-sequence item correlation are considered and encoded in two different modules. 
Figure 2 shows the workflow of the framework.
In particular, inter-sequence item correlation is depicted with graphs.
Graph neural networks are used to propagate the information along the paths between any two items.
We choose graph neural network for its ability in modeling high-order item correlation and for its promising performance in information propagation~\cite{NGCF}.

In summary, the main contributions are as follows:
\begin{itemize}
    \item We propose the \textbf{I}nter-\textbf{S}equence enhanced framework for personalized \textbf{S}equential \textbf{R}ecommendation (ISSR), which integrates both the intra-sequence and inter-sequence item correlation.
    \item In the inter-sequence item correlation encoder, graph neural networks are applied to encode high-order inter-sequence item correlation which is able to gather information from different sequences in an explicit manner.
    
    \item We conduct experimental studies on three real-world large-scale datasets. Extensive experimental results shows that 1) ISSR outperforms many state-of-the art models. 2) The performance of classic intra-sequence based models can be boosted significantly by adding up our proposed inter-sequence item correlation encoder. 3) The application of graph neural network in modeling inter-sequence correlation performs better than low-order, e.g. Matrix Factorization(MF)-based, methods.
\end{itemize}

\section{Related Work}

In this section, we review both the conventional and the deep learning-based methods for sequential recommendation.

\textbf{Conventional methods.}
Two categories of conventional methods can be applied for sequential recommendation.
The first category, such as Matrix Factorization~\cite{koren2009matrix} and k-nearest neighbor~\cite{KNN4Session} methods, relies on computing user-item or item-item similarities for recommendation. However, this line of works ignore the sequential patterns in users' behavior.
In the second category, such as shani et al.~\cite{shani2005mdp} tries to model item-item transitions in sequences with \emph{first-order} Markov chains to capture the sequential patterns. And the authors in~\cite{RendleFpmc} considers both user-item similarities and first-order item-item transitions for sequential recommendation (FPMC). For better capturing user's general interest and sequential patterns, Wang et al. \cite{HRM2015} extend FPMC by using a hierarchical structure to learn uses representation. Moreover, he et al.~\cite{he2016fusing} improve FPMC by utilizing \emph{high-order} Markov chains to solve the sparsity problem in sequential recommendation.
However, the above MC-based methods only model the intra-sequence interest between adjacent interactions.

\textbf{Deep learning-based methods.}
Recently, benefit from the powerful feature representation ability, deep learning-based methods are increasingly popular in sequential recommendation. Recurrent Neural Network is the most popular technique for sequential recommendation~\cite{HidasiKBT15Gru4rec,hidasi2018recurrent,DBLP:conf/www/RakkappanR19}, due to its inherent ability for modeling sequential dynamics. 
The authors in~\cite{HidasiKBT15Gru4rec} utilize Gated Recurrent Units (GRU) to model the sequential dynamics for session-based recommendation, and use session-parallel mini-batches technique to train the model. What's more, an improved version is proposed in~\cite{hidasi2018recurrent}, where a novel ranking loss function and an efficient sampling strategy are proposed.
In addition to the RNN-based methods, Convolutional Neural Network (CNN) is also adopted for sequential recommendation~\cite{Tangcaser,yuanfajiecnn,xu2019recurrent}. 
In~\cite{Tangcaser}, the researchers embeds the recent engaged items into an ``image" in the latent space, then employ different convolutional kernels to extract sequential patterns. 
And the authors in~\cite{yuanfajiecnn} improves the work in~\cite{Tangcaser} by applying dilated convolutional layers and residual block structure such that the performance can be improved, especially for long sequences. 
Moreover, Xu et al.~\cite{xu2019recurrent} combine RNN and CNN to learn user long- and short-term interest for sequential recommendation, where the hidden states of the RNN layer is the input the of the CNN layer.
However, such RNN- and CNN-based methods always encodes the user interactions into hidden states or latent factors without considering the different impacts of the items consumed at different time steps on current decision.
Therefore, the attention based models, which exhibit promising performance in sequence learning, are also utilized in sequential recommendation\cite{wangchengkangSelfattention,sun2019bert4rec}. In addition, the authors in~\cite{MaHGN} propose to utilize gated network for sequential recommendation, where a feature gating layer and an instance gating layer are employed to select what item features can be passed to the downstream layers from the feature and instance levels, respectively.
Recently, Graph Neural Networks (GNN) gains great attentions in recommendation community~\cite{perozzi2014deepwalk,tang2015line,grover2016node2vec,zhou2017scalable,shuwuSrgnn}, and the authors in~\cite{shuwuSrgnn} firstly employ GNN for session based recommendation. It models the sequences of items for each sessions separately, however, the ability of capturing the cross-sequence interest among the items is limited. 
Moreover, in other lines of work, transfer learning~\cite{DBLP:conf/sigir/MaRLCMR19} and variational model~\cite{DBLP:conf/www/XiaoLM19} are also utilized for sequential recommendation.

As we can see, the majority of the conventional and deep learning-based methods for sequential recommendation focus on intra-sequence interest within each individual sequence, but neglect the inter-sequence interest across different sequences. Different from above methods, our proposed MGSR coordinates the intra-sequence interest and inter-sequence interest modeling in sequential recommendation.

\section{Problem Formulation}


Let $\mathcal{U} = \{ {u_1},{u_2},...,{u_{|\mathcal{U}|}}\}$ and $\mathcal{V} = \{ {v_1},{v_2},...,{v_{|\mathcal{V}|}}\}$ denote the set of users and the set of items, respectively.
Given the sequences of interacted items from all users, the goal of \emph{sequential recommendation} is to recommend a list of items from $\mathcal{V}$ to each user $u_i\in \mathcal{U}$ such that the user is most likely to interact with the recommended items.
For a specific user, our framework outputs the probabilities for all candidate items, which represent how likely she will engage with the items based on her engaged sequence of items.


                                
\section{Methodology}

\begin{figure*}[!t]
\center
\includegraphics[width=0.75\textwidth]{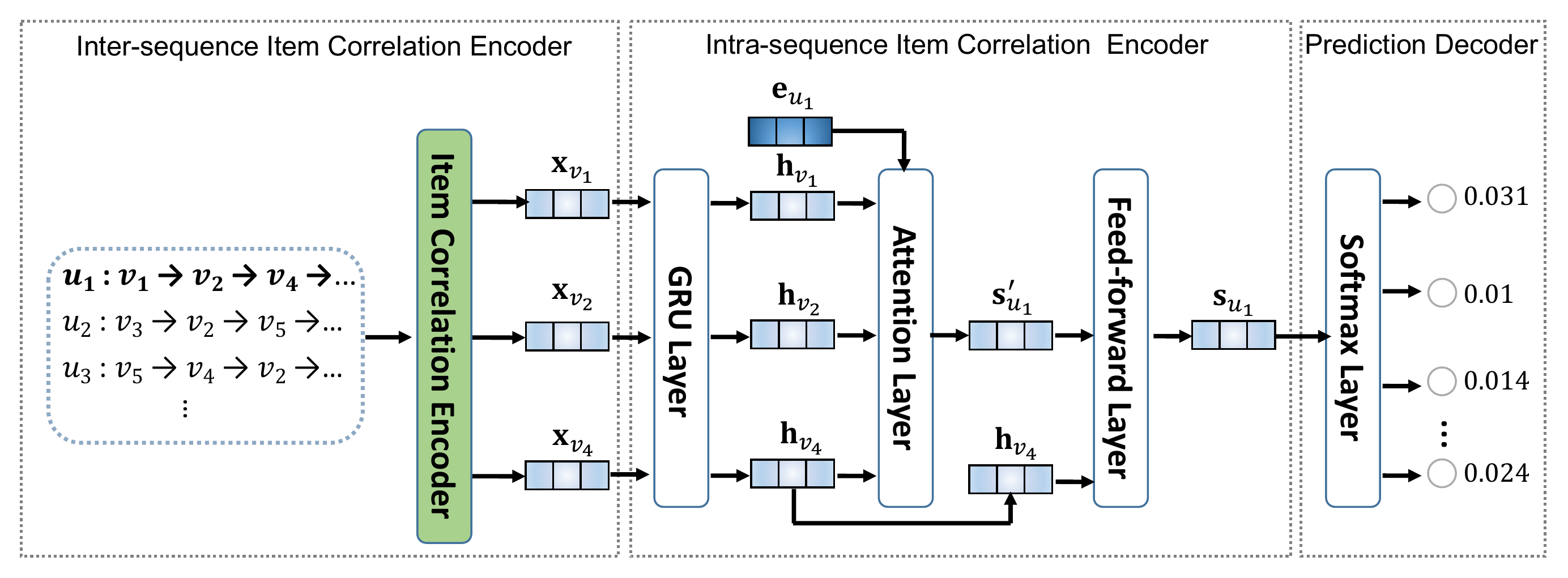}

\caption{\small{The Proposed Framework: ISSR.}}
\label{fig:framework}
\end{figure*}

As shown in Figure~\ref{fig:framework}, the proposed framework ISSR consists of an \emph{inter-sequence item correlation encoder}, an \emph{intra-sequence item correlation encoder}, and a \emph{prediction decoder}.
Firstly, we model the inter-sequence item correlation with two graphs, which are the user-item bipartite graph~\cite{NGCF} and the item-item co-occurrence graph~\cite{li2019graph}. Based on these graphs, the high-order inter-sequence item correlation can be captured through stacking multiple GNN layers.
Although the bipartite graph and co-occurrence graph have already been utilized in recommender systems, we are the first to combine the two graphs to exploit the high-order inter-sequence item correlation in sequential recommendation scenario.
Next, an intra-sequence item correlation encoder is developed, which pre-fuse the inter-sequence item correlation information with their intra-sequence sequential correlation and temporal dynamics. And we then integrate the item representations, which comprehensively captures both the inter- and intra-sequence item correlation, to generate the representation of the user's current interest.
Finally, in the prediction decoder, the user's preference on different items is computed based on the user's interest representation. Each part will be elaborated in the following.

\begin{figure}[!t]
\center
\includegraphics[width=0.48\textwidth]{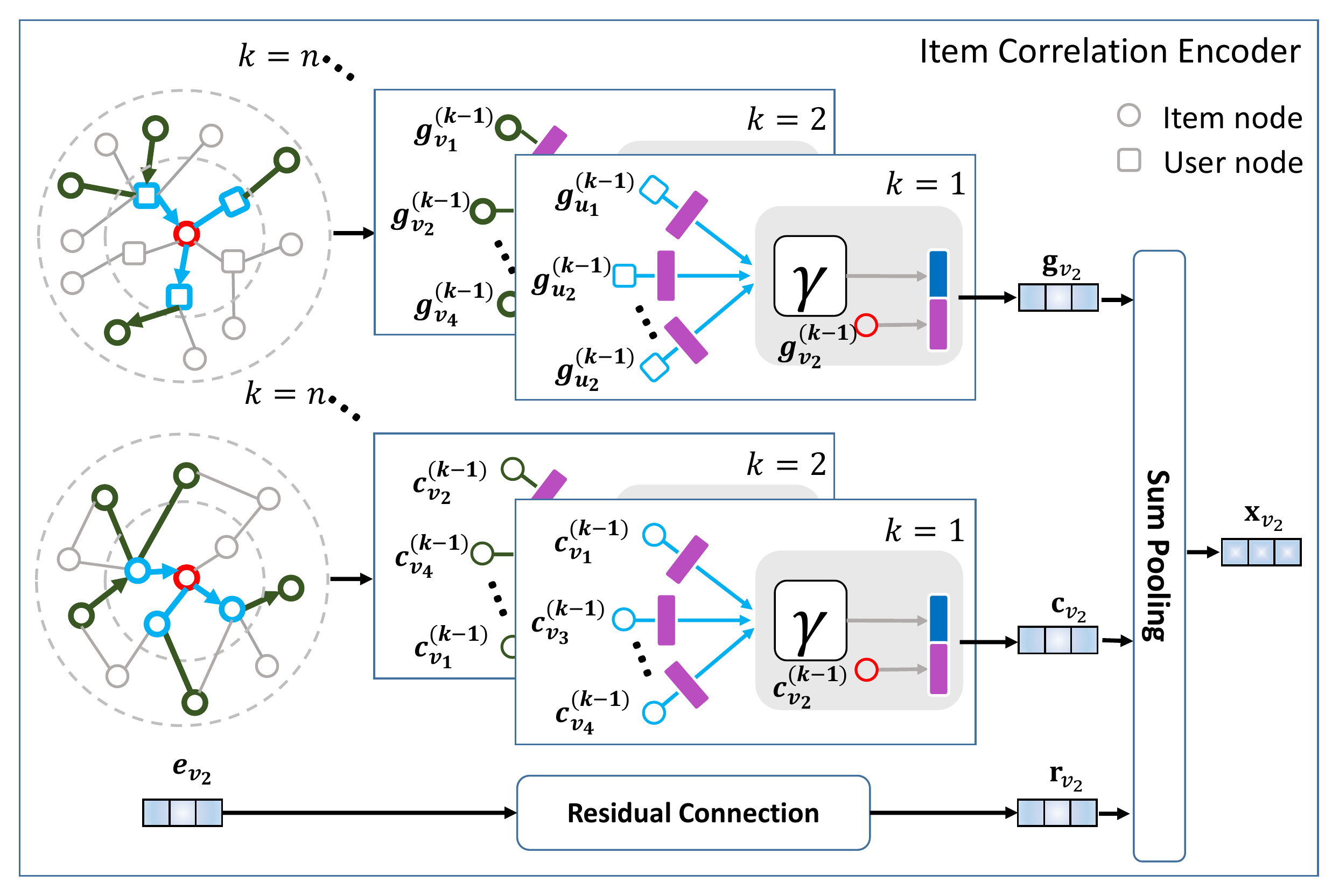}

\caption{\small{Item correlation encoder to capture the high-order inter-sequence item correlation.}}
\label{fig:framework1}
\end{figure}

\subsection{Inter-sequence Item Correlation Encoder}
To obtain the informative inter-sequence item correlation, we propose an item correlation encoder, which is specified in Figure \ref{fig:framework1}. 
As Figure \ref{fig:framework1} shows, ISSR exploits the item correlation from both the user-item bipartite graph and the item-item co-occurrence graph. 
In addition, ISSR also considers the residual connection to preserve the original item representation.
Finally, we generate the integrated item representation by fusing the three types of item information.

\subsubsection{Inter-sequence Item Correlation from User-Item Bipartite Graph.}\label{Method:Represent:Implicit}
The user-item bipartite graph contains two types of nodes, namely the user nodes and the item nodes. An edge exists between a user and an item if the user interacted with the item. For clarity, the adjacent nodes of a target node in the graph is defined as the $1$-hop neighbors of the target node. Particularly, for each node $v_i \in \mathcal{V}$ in the graph, $\mathcal{N}(v)$ denotes the set of 1-hop neighbors of $v_i$. 
Thus, the high-order inter-sequence item correlation can be seized via multiple hops on the graph through the user nodes.
For easy to follow, a path connected with multiple item nodes and user nodes is highlighted in the bipartite graph (the top left of Figure \ref{fig:framework1}), where the correlation between item nodes located in the start and the end of the path can be captured\footnote{Note that the arrows only highlight the paths, and the two graphs are undirected.}.
We apply graph convolutional network~\cite{rexyingPinsage} on the user-item bipartite graph (denoted as GCN$_B$) to aggregate the neighborhood information. As a result, the item correlation from the $k$-hop neighbors can be captured via stacking multiple GCN layers.





We denote the initial embedding of item node $v_i$ as $\mathbf{e}_{v_i}\in \mathbb{R}^{d}$ with dimension $d$ (or $\mathbf{e}_{u_i}\in \mathbb{R}^{d}$ for the user node $u$) and the hidden representation of $v_i$ at layer-$k$ as $\mathbf{g}_{v_i}^k \in \mathbb{R}^{d}$ (or as $\mathbf{g}_{u_i}^k$ for the user node $u$).
In GCN$_B$, a node embedding depends on both the node information and the graph structure around it. We first aggregate the neighborhood information of the target node (as shown in Eq. (\ref{equ:aggreation})), and then integrate the aggregated neighborhood information with the target node (as shown in Eq. (\ref{equ:transform})).

Specifically, we represent the neighborhood of an item node $v_i$ at layer-$k$ (or a user node $u_i$ at layer-$k$), by applying an aggregate function on all its neighbors at layer-($k$-$1$):

\begin{equation}\label{equ:aggreation}
\begin{split}
  \mathbf{z}_{v_i}^{k-1} = \sigma (\gamma \{ \mathbf{Q}_{\mathcal{V}}^k \cdot \mathbf{g}_{u_i}^{k - 1} + \mathbf{q}_{\mathcal{V}}^k | u \in \mathcal{N}(v)\}),~\mathbf{g}_{u_i}^0=\mathbf{e}_{u_i},\\
    \mathbf{z}_{u_i}^{k-1} = \sigma (\gamma \{ \mathbf{Q}_{\mathcal{U}}^k \cdot \mathbf{g}_{v_i}^{k - 1} + \mathbf{q}_{\mathcal{U}}^k | v \in \mathcal{N}(u)\}),~\mathbf{g}_{v_i}^0=\mathbf{e}_{v_i}.
 \end{split}
\end{equation}
$\mathbf{z}_{v_i}^{k-1}$ (or $\mathbf{z}_{u_i}^{k-1}$) is the representation of the neighborhood of item node $v_i$ (or of user node $u_i$) at layer-($k$-$1$). $\mathbf{Q}_{\mathcal{V}}^k \in \mathbb{R}^{d\times d}$ and $\mathbf{q}_{\mathcal{V}}^k \in \mathbb{R}^{d}$ (or, $\mathbf{Q}_{\mathcal{U}}^k$ and $\mathbf{q}_{\mathcal{U}}^k$) are the weight matrix and bias of the item (or user) aggregator at layer-$k$, respectively.
A pooling function $\gamma$ (such as weighted sum, weighted average, and etc.) is performed to aggregate the neighbor representations. $\sigma$ is an activation function. 

Note that different neural networks are utilized to transform the representations of the user nodes and item nodes from lower layers to higher layers, which differs form existing GCN works~\cite{rexyingPinsage,NGCF}, where normally a unified network is utilized. The reason is that user nodes and item nodes in bipartite graphs are intrinsically different and such difference should be considered when we aggregate information from user nodes or from item nodes. 

After generalizing the representation of the target node's neighborhood, we integrate such neighborhood representation with the current representation of the target node, as:

\begin{equation}\label{equ:transform}
\begin{aligned}
   & \mathbf{g}_{v_i}^k = \sigma (\mathbf{P}_{\mathcal{V}}^k \cdot [\mathbf{g}_{v_i}^{k - 1};\mathbf{z}_{v_i}^{k-1}] + \mathbf{p}_{\mathcal{V}}^k),\\
   & \mathbf{g}_{u_i}^k = \sigma (\mathbf{P}_{\mathcal{U}}^k \cdot [\mathbf{g}_{u_i}^{k - 1};\mathbf{z}_{u_i}^{k-1}] + \mathbf{p}_{\mathcal{U}}^k). \\
\end{aligned}
\end{equation}

\noindent $\mathbf{P}_{\mathcal{V}}^k \in \mathbb{R}^{d\times d}$ and $\mathbf{p}_{\mathcal{V}}^k \in \mathbb{R}^{d}$ (or, $\mathbf{P}_{\mathcal{U}}^k$ and $\mathbf{p}_{\mathcal{U}}^k$) are the transformation weight matrix and bias of the item (or user) at layer-$k$, respectively.
$[;]$ represents concatenation.





\subsubsection{Inter-sequence Item Correlation from Item-Item Co-occurrence Graph} 

The GCN$_B$ captures the item correlation through multiple-hop neighbors in the user-item bipartite graph. In addition, the item correlation can also be modeled from an item-item graph. Though there exists multiple methods to construct such an item-item graph. We utilize item co-occurrence information in users' behavior sequences to build it following~\cite{li2019graph}, which can be treated as a complementary of the user-item bipartite graph to directly model the frequency of item dependence. 

As shown in Figure~\ref{fig:framework1}, the co-occurrence graph includes a set of nodes where each node represents an item. 
An edge connects two nodes, if these two items are adjacent in a certain user's behavior sequence. The weight of an edge represents the number of times that the two items occur in users' behavior sequences, and such weight is utilized for neighborhood sampling, which we specify it later.
We apply the graph convolutional network on such item-item co-occurrence graph where we denote it as GCN$_C$. GCN$_C$ works differently from GCN$_B$ in the following two aspects. First, the item-item co-occurrence graph includes only one type of nodes, so that GCN$_C$ has neither aggregator weight matrices nor transformation weight matrices for user nodes. Second, the item-item co-occurrence graph associates weights on edges, which results in a different neighborhood sampling strategy as we will discuss in `Network Training' section. Due to the limit of space, we will not elaborate the details of GCN$_C$. We represent the embedding of an item node $v_i$ in this item-item co-occurrence graph as $\mathbf{c}_{v_i}$.

In summary, the user-item bipartite graph and the item-item co-occurrence graph are designed to capture the inter-sequence item correlation, which act as complementary roles.


\subsubsection{Residual Connection}
Inspired by~\cite{wangchengkangSelfattention}, we introduce the residual connection component in the inter-sequence item correlation encoder, which plays a role of preserving the original item representations: $\mathbf{r}_{v_i} = \sigma (\mathbf{W}_r \cdot \mathbf{e}_{v_i} + \mathbf{b}_r)$.
where $\mathbf{e}_{v_i}$ is the original embedding of the item $v_i$. The residual connection transforms the original embedding $\mathbf{e}_{v_i}$ to $\mathbf{r}_{v_i}$ via a hidden layer with $\mathbf{W}_r \in \mathbb{R}^{d\times d}$ and $\mathbf{b}_r \in \mathbb{R}^{d}$.

\subsubsection{Information Fusion}

As presented in Figure~\ref{fig:framework1}, the item representations learned from the two graphs and the residual connection are fused through an information fusion function $f(\cdot)$. More formally, the final representation $\mathbf{x}_{v_i}$ of an item $v_i$ is generated as: $\mathbf{x}_{v_i} = f(\mathbf{g}_{v_i},\mathbf{c}_{v_i},\mathbf{r}_{v_i}).$, and $f$ varies according to different application scenarios.
In this paper, we empirically choose \emph{element-wise sum} as the fusion function (sum pooling in Figure \ref{fig:framework1}) due to its superior performance compared to the other operations, such as concatenation and element-wise mean and gated networks~\cite{gate}.

\subsection{Intra-sequence Item Correlation Encoder}

In the intra-sequence item correlation encoder, ISSR aims to model the intra-sequence item correlation with considering the inter-sequence item correlation captured in the inter-sequence item correlation encoder, and finally integrate the user's current interest over the candidate items.
As presented in Figure~\ref{fig:framework}, a GRU layer is performed to capture the intra-sequence item sequential correlation among the items in a sequence, where the hidden states in the GRU layer represent user's interests at different time steps.
Then an attention network is utilized to aggregate the user's interests at different time steps, generating the final representation of the user's current interest.

\subsubsection{The GRU Layer}

As shown in Figure \ref{fig:framework}, the input of the GRU layer is a sequence of items, of which the embeddings are learned from the inter-sequence item correlation encoder.
The GRU layer outputs a sequence of hidden vectors $\{\mathbf{h}_{v_i}\}$, which represent the user's interests at different time steps.

\subsubsection{The Attention Layer}

User's interests at different time steps contribute differently to the user's current decision.
Moreover, different users have different levels of sensitivity to the temporal dynamics.
Due to such motivations, we devise a \emph{personalized} attention network to capture the evolving interests of each particular user.

Specifically, the inputs of the attention network are the sequence of hidden states generated by the GRU layer, e.g., $({\mathbf{h}_{v_1}},{\mathbf{h}_{v_2}},...,{\mathbf{h}_{v_L}})$, and the embedding of the user $\mathbf{e}_{u_i}$. 
With multi-layer perceptrons, the attention network generates a weight for each input hidden state  representing the contribution of the user's interest at that time step (represented by the input hidden vector) on the user's final decision.

\begin{equation}\label{equ:attentionnet}
\begin{split}
& {a'_{{u_i}{v_j}}} = \mathbf{W}_1 \sigma(\mathbf{W}_2 ( [\mathbf{e}_{u_i}; \mathbf{h}_{v_j}]) + \mathbf{b}_2) + \mathbf{b}_1 ,\\
& {a_{{u_i}{v_j}}} = \frac{\exp(a'_{{u_i}{v_j}})}  {\sum_{1\leq t \leq L } \exp(a'_{{u_i}{v_t}}) } .
\end{split}
\end{equation}

As shown in Eq. (\ref{equ:attentionnet}), ${a_{{u_i}{v_j}}}$ is the weight of the hidden state $\mathbf{h}_{v_j}$ on the final decision of user $u_i$.
$\mathbf{W}_1 \in \mathbb{R}^{d\times d}, \mathbf{W}_2\in \mathbb{R}^{d\times d}, \mathbf{b}_1\in \mathbb{R}^{d}, \mathbf{b}_2\in \mathbb{R}^{d}$ are parameters of the multi-layer perceptrons. 
For simplicity, we only present two layers of parameters in multi-layer perceptions in Eq. (\ref{equ:attentionnet}), which is also the case in our implementation.
The user's interest, considering both the intra-sequence item sequential correlation and the temporal dynamics, is computed as:

\begin{equation}\label{equ:localinterest}
{\mathbf{s}'_u} = \sum_{1\leq j \leq L } {{a_{{u_i}{v_j}}}{\mathbf{h}_{v_j}}}.
\end{equation}

As shown in~\cite{shuwuSrgnn}, the latest engaged item is able to reflect user's most recent interest. 
Therefore, we integrate the latest hidden state into user's final interest representation: $\mathbf{s}_{u} = \mathbf{W}_h [\mathbf{s}'_u;\mathbf{h}_L].$ 



\subsection{Prediction Decoder}

After obtaining the representation of the user's current interest,
we adopt the classic matrix factorization approach to infer the user's preference on the items.
The prediction score of user $u_i$ on item $v_i$ is the inner product of the user's interest $\mathbf{s}_{u_i}$ and the item embedding $\mathbf{e}_{v_i}$.
The probability that the user will interact with the item is defined as the \textit{softmax} of the prediction score: $\hat y_i = \text{softmax}({\mathbf{s}_{u_i} ^\top} \mathbf{e}_{v_i})$.


\subsection{Network Training}
\subsubsection{Loss Function}
To generate the training data, we extract $L$ consecutive items in a sequence as user's behavior sequence and the following $T$ items as positive samples. 
We also sample a number of items that the user did not interacted with as negative samples following~\cite{wangchengkangSelfattention}.
We adopt \textit{cross entropy} as the training loss to describe the discrepancy between the predicted probabilities and the ground truth labels as shown in Eq. (\ref{equ:loss}) .

\begin{equation}\label{equ:loss}
\mathcal{L}({{\hat y}_i}) =  - \sum\limits_{i = 1}^{n} {{y_i}\log ({{\hat y}_i}) + (1 - {y_i})\log (1 - {{\hat y}_i})},
\end{equation}
\noindent where $n$ is the number of instances,  $y_i=1$ if the predicted item is engaged by the user; otherwise, $y_i=0$.

\subsubsection{Neighborhood Sampling}

We adopt neighborhood sampling techniques to facilitate the network training.
Specifically, for GCN$_{B}$ on the bipartite graph, we sample $10$ neighbors for each node uniformly at random.
For GCN$_{C}$ on the co-occurrence graph, we apply importance sampling, which samples $10$ neighbors for each node according to the weights of the edges.


\section{Experiments}
In this section, we compare the proposed framework with the state-of-the-art methods on three real-world datasets.
We also comprehensively analyze the results of the proposed framework under different experimental settings.


\subsection{Datasets}


  
  

Experiments are conducted on the following three public benchmark datasets: MovieLens (1M) (abbreviated as ML (1M)), Steam~\cite{wangchengkangSelfattention}, MovieLens (20M) (abbreviated as ML (20M)), which is of different scales and sparsity.
We process the three datasets following the existing research~\cite{Tangcaser}, in which all the ratings are treated as implicit feedbacks. 
And the items in a user's sequence are sorted in chronological order.
We hold the first 70\%, following 10\%  and last 20\% of items in each user's sequence as the \emph{training set}, the \emph{validation set} and the \emph{testing set}, respectively\footnote{Note that the data partition strategy is different from the original SASRec~\cite{wangchengkangSelfattention}, they treat the last item in the whole sequence of each user as the test set, the second last item as validation set, and all the previous items as training set. So their training set is much larger than the one utilized in this paper while the test set is much smaller. 
Moreover, in HGN, the authors treat all the ratings less than four as negative samples, and then filter the noise data.  However, in Caser, the authors treat all the ratings as implicit feedbacks. Therefore, the size of the ML (20) dataset in HGN is only half of that in this paper.}.
The statistics of the three processed datasets are summarized in Table~\ref{tab:dataset}.



\begin{table}[ht]
  \centering 

  \caption{\small{Statistics of the Datasets.}}

  \resizebox{0.46\textwidth}{!}{
  \begin{tabular}{l|ccccc}
    \toprule
    {Datasets}    & {\#users} & {\#items} & avg.\#items/user & {sparsity} &{\#interactions}\\
    \midrule
  ML (1M)        & 6,040   & 3,416    & 165.50  & 95.16\%   & 999,611\\
  Steam        & 334,730 & 13,047   & 11.01   & 99.92\%   & 3,686,172\\
  ML (20M)         & 138,493 & 15,451   & 144.16  & 99.07\%   & 19,964,833\\
  \bottomrule
  \end{tabular}
  }
  \label{tab:dataset}
\end{table}

\subsection{Experimental Settings}

\subsubsection{Evaluation Metrics}

Following~\cite{Tangcaser}, we adopt \textit{Recall@$k$}, \textit{nDCG@$k$}, \textit{HR@$k$} and \textit{MRR@$k$} for $k\in\{5,10\}$ to evaluate the effectiveness of different methods.



\subsubsection{Compared Methods}

To demonstrate the superiority of the proposed ISSR, we compare it to several state-of-the-art baselines. 
For easy to follow, we summarize them into two categories according to whether they can model the inter- and intra-sequence item correlation, namely (1) \textbf{only intra-sequence based methods}: \textbf{GRU4Rec}~\cite{HidasiKBT15Gru4rec}, \textbf{Caser}~\cite{Tangcaser}, \textbf{SASRec}~\cite{wangchengkangSelfattention}, \textbf{SR-GNN}~\cite{shuwuSrgnn} and \textbf{HGN}~\cite{MaHGN}; (2) \textbf{both inter- and intra-sequence based methods}: \textbf{FPMC}~\cite{RendleFpmc}, \textbf{Fossil}~\cite{he2016fusing} and \textbf{CSRM}~\cite{DBLP:conf/sigir/WangRMCMR19}.

Methods such as the popularity~\cite{cremonesi2010performance} and BPR-MF~\cite{rendle2009bpr} which we treat as \textbf{only inter-sequence based methods} are omitted in comparison, because they are proven to be inferior to the compared methods as they lack the ability of modeling \textbf{intra-sequence} sequential correlation~\cite{Tangcaser,MaHGN}.

We implement FPMC, Fossil and ISSR using TensorFlow with Adam~\cite{kingma2014adam} optimizer.
We utilize the source code of GRU4Rec, Caser, SASRec, SR-GNN, CSRM, HGN
to re-produce their performance. 

\begin{table*}[t!]
  \centering
  \large
  \caption{\small{Overall Performance Comparison ($\star$ denotes statistically significant improvement with p-value$<1e-5$).}}

  \label{table: overall result}
  \resizebox{0.9\textwidth}{!}{

  \begin{tabular}{c|c|c|cccccccc}
  \toprule

  Dataset                   & \multicolumn{2}{c|}{Model}        & Recall@5  & Recall@10   & nDCG@5 & nDCG@10  & HR@5    & HR@10    & MRR@5  & MRR@10  \\
  \hline                                
  \multirow{10}{*}{ML (1M)} &\multirow{5}{*}{Intra}   
                                                     & GRU4Rec      & 0.0707    & 0.1232  & 0.4661  & 0.5012  & 0.1558  & 0.2507  & 0.0792  & 0.0918  \\
                            &                        & Caser        & 0.0764    & 0.1326  & 0.4783  & 0.5176  & 0.1613  & 0.2624  & 0.0826  & 0.0961  \\
                            &                        & SASRec       & 0.0812    & 0.1320  & 0.4778  & 0.5156  & 0.1726  & 0.2773  & 0.0891  & 0.1018  \\
                            &                        & SR-GNN     &\underline{0.0834}&\underline{0.1385}&0.4859&0.5234&\underline{0.1854}&0.2906&\underline{0.1034}&0.1152 \\
                            &                        & HGN         & 0.0816    & 0.1378  &\underline{0.4973}&\underline{0.5295}& 0.1836  &\underline{0.2945}& 0.1012 & \underline{0.1154}\\
                            \cline{2-11}

                            & intra                  & FPMC        & 0.0653    & 0.1095  & 0.4183  & 0.4632  & 0.1567  & 0.2554  & 0.0798  & 0.0904 \\ 
                            & and                    & Fossil      & 0.0701    & 0.1213  & 0.4555  & 0.4925  & 0.1611  & 0.2501  & 0.0803  & 0.0912 \\
                            & inter                  & CSRM        & 0.0815    & 0.1381  & 0.4901  & 0.5238  & 0.1768  & 0.2836  & 0.0965  & 0.1107 \\
                            &                        & ISSR          &$\mathbf{0.0934^{\star}}$ &$\mathbf{0.1563^{\star}}$ &$\mathbf{0.5354^{\star}}$ &$\mathbf{0.5632^{\star}}$ &$\mathbf{0.2157^{\star}}$ &$\mathbf{0.3318^{\star}}$ &$\mathbf{0.1186^{\star}}$ &$\mathbf{0.1337^{\star}}$ \\
                            
                            \cline{2-11}

                            & \multicolumn{2}{c|}{$\%$ Improv.}    &11.99\% & 12.85\% & 7.66\%  & 6.36\%  & 16.34\% & 12.67\% & 14.70\% & 15.86\%  \\

  \hline
  \multirow{8}{*}{Steam}    &\multirow{5}{*}{Intra}    
                                                     & GRU4Rec    & 0.0381  & 0.0714  & 0.0485  & 0.0702  & 0.0404  & 0.0755  & 0.0134  & 0.0181 \\
                            &                        & Caser      & 0.0408  & 0.0744  & 0.0507  & 0.0735  & 0.0459  & 0.0825  & 0.0158  & 0.0206\\
                            &                        & SASRec     & 0.0403  & 0.0737  & 0.0502  & 0.0726  & 0.0447  & 0.0821  & 0.0151  & 0.0199 \\
                            &                        & SR-GNN     & 0.0415  & 0.0759  & 0.0548  & 0.0782  & 0.0472  & 0.0837  & 0.0181  & 0.0232 \\
                            &                        & HGN        &\underline{0.0437}&\underline{0.0767}&\underline{0.0588}&\underline{0.0803}&\underline{0.0495}&\underline{0.0851}&\underline{0.0200}&\underline{0.0247}\\
                            \cline{2-11}
                            
                            &intra                   & FPMC       & 0.0354  & 0.0612  & 0.0491  & 0.0663  & 0.0380  & 0.0658  & 0.0174  & 0.0210 \\ 
                            &and                     & Fossil     & 0.0408  & 0.0745  & 0.0504  & 0.0721  & 0.0455  & 0.0800  & 0.0171  & 0.0232 \\
                            &inter                   & CSRM       & 0.0427  & 0.0747  & 0.0573  & 0.0773  & 0.0453  & 0.0825  & 0.0197  & 0.0239 \\
                            &                        & ISSR       &$\mathbf{0.0492^{\star}}$ &$\mathbf{0.0872^{\star}}$ &$\mathbf{0.0644^{\star}}$ &$\mathbf{0.0884^{\star}}$ &$\mathbf{0.0551^{\star}}$ &$\mathbf{0.0972^{\star}}$ &$\mathbf{0.0225^{\star}}$ &$\mathbf{0.0277^{\star}}$ \\
                            \cline{2-11}

                            & \multicolumn{2}{c|}{$\%$ Improv.}   & 12.59\% & 13.69\% & 9.52\%  & 10.09\% & 11.31\% & 14.22\% & 11.94\% & 12.15\% \\
  \hline
  \multirow{8}{*}{ML (20M)} &\multirow{5}{*}{Intra}   
                                                    & GRU4Rec     & 0.0584  & 0.1031  & 0.3171  & 0.3612  & 0.0963  & 0.1630  & 0.0477  & 0.0564 \\
                            &                       & Caser       & 0.0644  & 0.1127  & 0.3371  & 0.3829  & 0.1045  & 0.1750  & 0.0530  & 0.0623 \\
                            &                       & SASRec      & 0.0624  & 0.1102  & 0.3341  & 0.3794  & 0.1177  & 0.1940  & 0.0609  & 0.0756\\
                            &                       & SR-GNN      & 0.0752  & 0.1271  & 0.3643  & 0.4060  & 0.1349  & 0.2142  & 0.0710  & 0.0814\\
                  
                            &                       & HGN         &\underline{0.0812}&\underline{0.1350}&\underline{0.3749}&\underline{0.4135}&\underline{0.1468}&\underline{0.2288}&\underline{0.0780}&\underline{0.0888} \\ 
                            \cline{2-11}
                            
                            &intra                   & FPMC      & 0.0569  & 0.0982  & 0.3054  & 0.3459  & 0.0928  & 0.1515  & 0.0476  & 0.0548 \\ 
                            & and                    & Fossil    & 0.0607  & 0.1030  & 0.3152  & 0.3681  & 0.0989  & 0.1589  & 0.0483  & 0.0562 \\
                            &inter                   & CSRM      & 0.0724  & 0.1259  & 0.3625  & 0.4054  & 0.1370  & 0.2102  & 0.0650  & 0.0759 \\
                            &                        & ISSR      &$\mathbf{0.0883^{\star}}$ &$\mathbf{0.1485^{\star}}$ &$\mathbf{0.4073^{\star}}$ &$\mathbf{0.4452^{\star}}$ &$\mathbf{0.1619^{\star}}$ &$\mathbf{0.2569^{\star}}$ &$\mathbf{0.0848^{\star}}$ &$\mathbf{0.0973^{\star}}$ \\
                            \cline{2-11}

                            & \multicolumn{2}{c|}{$\%$ Improv.}  & 8.74\% & 10.00\% & 8.64\%  & 7.67\%  & 10.29\% & 12.28\% & 8.72\%  & 9.57\% \\
  \bottomrule 
  \end{tabular}
  }
\end{table*}

\subsubsection{Parameter Settings}

The best hyper-parameters for each model are found from exhaustive search on the validation set.
In particular, for Caser, the number of the vertical and horizontal filters are searched from $\{1, 2, 4, 8, 16, 32, 64\}$.
For SASRec, the number of self-attention blocks is searched from $\{1, 2, 3\}$.
For SR-GNN and HGN, we follow the best parameter settings in the original paper.
For CSRM, the memory size is searched from $\{128, 256, 512\}$ for different dataset. 
For our proposed framework ISSR, the neighborhood information is aggregated from at most 3-hop due to the efficiency concern. The best performance is observed when we consider 2-hop neighbors in GCN$_B$ and 1-hop neighbors in GCN$_C$.
We follow the settings in~\cite{Tangcaser,MaHGN} to set the sequence length to be 5 (i.e., $L=5$) and the number of subsequent items to be 3 (i.e., $T=3$) for all methods, unless stated otherwise, for fair comparison.






\subsection{Overall Performance Comparison}

Table~\ref{table: overall result} summarizes the overall performance of all the compared models on the three datasets, where the underlined numbers are the best results of the baselines, and the bold numbers are the best results of all models. $\star$ indicates the statistically significant improvement~\cite{ruxton2006unequal} (i.e., two-sided t-test) with p-value$<1e-5$ over the best baselines.
The row ``\%Improv." indicates the relative improvement of ISSR compared to the best baselines. We have the following observations. 

ISSR is more superior than HGN, SR-GNN, SASRec, Caser and GRU4Rec. This is due to three possible reasons.
\emph{First}, except for capturing the intra-sequence item correlation with GRU and attention layers, the two graphs constructed by ISSR capture the inter-sequence item correlation across different users.
Differently, no matter how the session graph is constructed by SR-GNN or the other methodology utilized in baselines, they only consider the intra-sequence item correlation within each individual session.
\emph{Second}, residual connections in ISSR preserve the original item representations, served as a complementary to the item representations produced by the graph neural network, which makes the item representations more informative.
\emph{Third}, the attention network in ISSR is personalized based on the fact that users may have different levels of sensitivity to temporal dynamics, which is not considered by the baseline methods.

ISSR outperforms FPMC, Fossil and CSRM, the reasons are as follows: (1) for FPMC and Fossil, ISSR captures high-order inter-sequence item correlation from the two graph neural networks with multiple hops, whereas, FPMC and Fossil only capture low-order inter-sequence item correlation with MF and similarity based model; In addition, FPMC and Fossil both utilize Markov Chains to model the intra-sequence item sequential correlation, which is proven to be inferior to deep neural network based model~\cite{HidasiKBT15Gru4rec,Tangcaser,wangchengkangSelfattention}; (2) for CSRM, it captures inter-sequence correlation with a simple session-level nearest neighbor-based approach instead of the fine-grained item-level correlation we addressed in ISSR.

Another observation is that, on the sparse Steam dataset, Fossil achieves comparable performance to Caser and SASRec, or even slightly better w.r.t. Recall@10 etc. evaluation metrics. This observation also indicates that explicitly modeling the inter-sequence item correlation can remedy the data scarce issue to some extent.

To summarize, ISSR shows consistent and significant improvement over the compared baselines on all the (dense or sparse, small or large scale) conducted datasets in terms of Recall, nDCG, HR and MRR, which demonstrates the superiority of our proposed framework. 

\subsection{Effect of Inter-sequence Item Correlation Encoder for Sequential Recommendation}
To verify the effectiveness of the proposed inter-sequence item correlation encoder, we incorporate the proposed inter-sequence item correlation encoder into the existing RNN-based, CNN-based and attention-based sequential recommendation methods, respectively. The results are reported in Table \ref{tab:ablation of inter}. From Table \ref{tab:ablation of inter}, we observe that these models are enhanced after equipping the proposed inter-sequence item correlation encoder. For instance, we achieve 25.0\%, 14.8\% and 9.0\% improvements in terms of Recall@10 on MovieLens (1M) datasets. The observation demonstrates that modeling the inter-sequence item correlation as in the proposed ISSR can boost the performance of existing SR methods.

\begin{table}[!t]
  \centering

  \caption{\small{Effect of inter-sequence item correlation module on ML (1M) and Steam datasets.}}

  \resizebox{0.45\textwidth}{!}{
  \begin{tabular}{l|cc|cc}
    \toprule
    \multirow{2}{*}{variants}    & \multicolumn{2}{c|}{ML(1M)} & \multicolumn{2}{c}{Steam} \\
    \cline{2-5}
                                 & Recall@10   & nDCG@10    & Recall@10    & nDCG@10  \\
    \midrule
    
  GRU4Rec                 & 0.1232      & 0.5012     & 0.0714       & 0.0702  \\
  Inter+GRU4Rec               & 0.1545      & 0.5609     & 0.0856       & 0.0872  \\ 
  \midrule
  SASRec                  & 0.1320      & 0.5156     & 0.0737       & 0.0726  \\
  Inter+SASRec               & 0.1558      & 0.5621     & 0.0864       & 0.0878  \\
  \midrule
  Caser                   & 0.1326      & 0.5176     & 0.0744       & 0.0735  \\
  Inter+Caser             & 0.1445      & 0.5460     & 0.0815       & 0.0828  \\
  \bottomrule
  \end{tabular}
  }
  \label{tab:ablation of inter}
\end{table}

\subsection{Effect of the Graphs in Inter-Sequence Item Correlation Encoder}
To verify the effectiveness of the bipartite and the co-occurrence graphs in capturing high-order inter-sequence item correlations, we design several variants of the inter-sequence encoder. These variants are \emph{Only intra}, \emph{MF+intra}, \emph{Co+intra}, \emph{Bi+intra} and ISSR as shown in Tables \ref{tab:ablation of GCN_B and GCN_C} where ISSR is our proposed framework. 
They represent null, MF-based low order, co-occurrence graph based high order, bipartite graph based high order and dual graph based high order inter-sequence encoder, respectively.
From Table \ref{tab:ablation of GCN_B and GCN_C}, we have the following observations. (1) \emph{Only intra} performs obviously worse than the other variants which demonstrates the indispensability of the inter-sequence encoder.
(2) The performance of \emph{MF+intra} is inferior to the graph based inter-sequence encoder due to MF's inability in explicitly capturing higher order item correlations across different sequences. Graph neural networks can model the high-order item correlations across different sequences with multiple hops.
(3) The minor difference observed between \emph{Bi+intra} and \emph{Co+intra} indicates user-item bipartite graph and item-item co-occurrence graph contribute almost equally to the final performance, and combining the two graphs together will enhance the performance.

\begin{table}[!t]
  \centering
  \caption{\small{Effect of the bipartite and the co-occurrence GCNs on ML (1M) and Steam.} }

  \resizebox{0.45\textwidth}{!}{
  \begin{tabular}{l|cc|cc}
    \toprule
    \multirow{2}{*}{variants}    & \multicolumn{2}{c|}{ML(1M)} & \multicolumn{2}{c}{Steam} \\
    \cline{2-5}
                                 & Recall@10   & nDCG@10    & Recall@10    & nDCG@10  \\
    \midrule
    Only intra              & 0.1378      & 0.5244     & 0.0747       & 0.0772  \\
    MF+intra                & 0.1412      & 0.5265     & 0.0766       & 0.0787  \\
    Co+intra                & 0.1541      & 0.5604     & 0.0852       & 0.0865  \\
    Bi+intra                & 0.1548      & 0.5608     & 0.0856       & 0.0869  \\
    ISSR                    & 0.1563      & 0.5632     & 0.0872       & 0.0884  \\

  \bottomrule
  \end{tabular}
  }
  \label{tab:ablation of GCN_B and GCN_C}
\end{table}

\subsection{Effect of Dimensionality of Item Embedding}

\begin{figure}[!ht]
\center
  \includegraphics[width=0.475\textwidth]{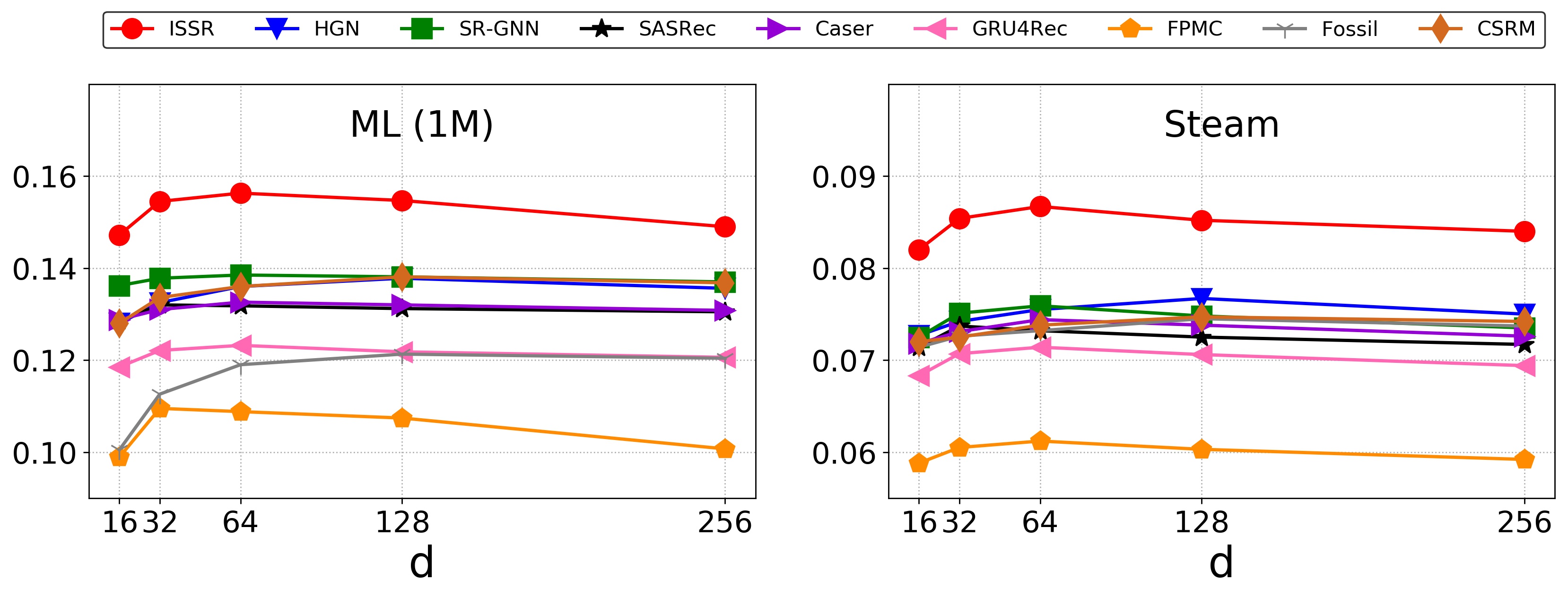}
\caption{\small{Effect of the Dimension $d$ on Recall@10.}}
\label{fig:d}
\end{figure}

Figure \ref{fig:d} presents the performance of all the compared models varying the dimension $d$ of item embedding. Due to the space limit, we only present the Recall@10 on ML(1M) and Steam.
As we increase $d$, the performance of all the models increases until reaching the best values.
Then, the performance drops or keeps stable as we continue increasing $d$.
The reason is that the capacity of the models increases as the dimensionality of item embeddings is enlarged. However, after reaching its peak value, the model capacity will not keep increasing even if the dimensionality of item embeddings continues increasing since the model capacity is limited by the amount of informative data.
ISSR reaches its best performance when setting $d=64$.

\section{Conclusion}
In this paper, we propose a \textbf{I}nter-\textbf{S}equence enhanced framework for personalized \textbf{S}equential \textbf{R}ecommendation (ISSR).
The inter-sequence item correlation encoder in ISSR utilizes two graphs (i.e., a user-item bipartite graph and an item-item co-occurrence graph) to capture the inter-sequence item correlations.
The intra-sequence item correlation encoder aggregates the learned inter-sequence item correlation information and considers the item sequential correlations and temporal dynamics in a current sequence to generate the representation of users' interests.
Then the user's next behavior on the candidate items can be predicted by the learned interests.
Extensive experiments on three real-world datasets are conducted. The results demonstrate the superiority of ISSR over many state-of-the-art methods.

\newpage

\bibliographystyle{aaai}
\bibliography{ref}

\begin{thebibliography}{}

\bibitem[\protect\citeauthoryear{Cremonesi, Koren, and
  Turrin}{2010}]{cremonesi2010performance}
Cremonesi, P.; Koren, Y.; and Turrin, R.
\newblock 2010.
\newblock Performance of recommender algorithms on top-n recommendation tasks.
\newblock In {\em Proceedings of the fourth ACM conference on Recommender
  systems, RecSys},  39--46.
\newblock ACM.

\bibitem[\protect\citeauthoryear{Dauphin \bgroup et al\mbox.\egroup
  }{2017}]{gate}
Dauphin, Y.~N.; Fan, A.; Auli, M.; and Grangier, D.
\newblock 2017.
\newblock Language modeling with gated convolutional networks.
\newblock In {\em ICML},  933--941.
\newblock JMLR. org.

\bibitem[\protect\citeauthoryear{Grover and
  Leskovec}{2016}]{grover2016node2vec}
Grover, A., and Leskovec, J.
\newblock 2016.
\newblock node2vec: Scalable feature learning for networks.
\newblock In {\em Proceedings of the 22nd ACM SIGKDD international conference
  on Knowledge discovery and data mining},  855--864.
\newblock ACM.

\bibitem[\protect\citeauthoryear{Guo \bgroup et al\mbox.\egroup
  }{2019}]{KNN4Session}
Guo, H.; Tang, R.; Ye, Y.; Liu, F.; and Zhang, Y.
\newblock 2019.
\newblock A novel {KNN} approach for session-based recommendation.
\newblock In {\em PAKDD},  381--393.

\bibitem[\protect\citeauthoryear{He and McAuley}{2016}]{he2016fusing}
He, R., and McAuley, J.
\newblock 2016.
\newblock Fusing similarity models with markov chains for sparse sequential
  recommendation.
\newblock In {\em ICDM},  191--200.
\newblock IEEE.

\bibitem[\protect\citeauthoryear{Hidasi and
  Karatzoglou}{2018}]{hidasi2018recurrent}
Hidasi, B., and Karatzoglou, A.
\newblock 2018.
\newblock Recurrent neural networks with top-k gains for session-based
  recommendations.
\newblock In {\em CIKM},  843--852.
\newblock ACM.

\bibitem[\protect\citeauthoryear{Hidasi \bgroup et al\mbox.\egroup
  }{2016}]{HidasiKBT15Gru4rec}
Hidasi, B.; Karatzoglou, A.; Baltrunas, L.; and Tikk, D.
\newblock 2016.
\newblock Session-based recommendations with recurrent neural networks.
\newblock In {\em 4th International Conference on Learning Representations,
  {ICLR} 2016, San Juan, Puerto Rico, May 2-4, 2016, Conference Track
  Proceedings}.

\bibitem[\protect\citeauthoryear{Kang and
  McAuley}{2018}]{wangchengkangSelfattention}
Kang, W., and McAuley, J.~J.
\newblock 2018.
\newblock Self-attentive sequential recommendation.
\newblock In {\em {IEEE} International Conference on Data Mining, {ICDM} 2018,
  Singapore, November 17-20, 2018},  197--206.

\bibitem[\protect\citeauthoryear{Kingma and Ba}{2014}]{kingma2014adam}
Kingma, D.~P., and Ba, J.
\newblock 2014.
\newblock Adam: A method for stochastic optimization.
\newblock {\em arXiv preprint arXiv:1412.6980}.

\bibitem[\protect\citeauthoryear{Koren, Bell, and
  Volinsky}{2009}]{koren2009matrix}
Koren, Y.; Bell, R.; and Volinsky, C.
\newblock 2009.
\newblock Matrix factorization techniques for recommender systems.
\newblock {\em Computer} (8):30--37.

\bibitem[\protect\citeauthoryear{Li \bgroup et al\mbox.\egroup
  }{2019}]{li2019graph}
Li, F.; Chen, Z.; Wang, P.; Ren, Y.; Zhang, D.; and Zhu, X.
\newblock 2019.
\newblock Graph intention network for click-through rate prediction in
  sponsored search.
\newblock In {\em SIGIR},  961--964.
\newblock ACM.

\bibitem[\protect\citeauthoryear{Ma \bgroup et al\mbox.\egroup
  }{2019}]{DBLP:conf/sigir/MaRLCMR19}
Ma, M.; Ren, P.; Lin, Y.; Chen, Z.; Ma, J.; and de~Rijke, M.
\newblock 2019.
\newblock {\(\pi\)}-net: {A} parallel information-sharing network for
  shared-account cross-domain sequential recommendations.
\newblock In {\em Proceedings of the 42nd International {ACM} {SIGIR}
  Conference on Research and Development in Information Retrieval, {SIGIR}
  2019, Paris, France, July 21-25, 2019},  685--694.

\bibitem[\protect\citeauthoryear{Ma, Kang, and Liu}{2019}]{MaHGN}
Ma, C.; Kang, P.; and Liu, X.
\newblock 2019.
\newblock Hierarchical gating networks for sequential recommendation.
\newblock In {\em KDD 2019, Anchorage, AK, USA, August 4-8, 2019},  825--833.

\bibitem[\protect\citeauthoryear{Perozzi, Al-Rfou, and
  Skiena}{2014}]{perozzi2014deepwalk}
Perozzi, B.; Al-Rfou, R.; and Skiena, S.
\newblock 2014.
\newblock Deepwalk: Online learning of social representations.
\newblock In {\em Proceedings of the 20th ACM SIGKDD international conference
  on Knowledge discovery and data mining},  701--710.
\newblock ACM.

\bibitem[\protect\citeauthoryear{Quadrana \bgroup et al\mbox.\egroup
  }{2017}]{DBLP:conf/recsys/QuadranaKHC17}
Quadrana, M.; Karatzoglou, A.; Hidasi, B.; and Cremonesi, P.
\newblock 2017.
\newblock Personalizing session-based recommendations with hierarchical
  recurrent neural networks.
\newblock In {\em Proceedings of the Eleventh {ACM} Conference on Recommender
  Systems, RecSys 2017, Como, Italy, August 27-31, 2017},  130--137.

\bibitem[\protect\citeauthoryear{Rakkappan and
  Rajan}{2019}]{DBLP:conf/www/RakkappanR19}
Rakkappan, L., and Rajan, V.
\newblock 2019.
\newblock Context-aware sequential recommendations withstacked recurrent neural
  networks.
\newblock In {\em {WWW} 2019, San Francisco, CA, USA, May 13-17, 2019},
  3172--3178.

\bibitem[\protect\citeauthoryear{Rendle \bgroup et al\mbox.\egroup
  }{2009}]{rendle2009bpr}
Rendle, S.; Freudenthaler, C.; Gantner, Z.; and Schmidt-Thieme, L.
\newblock 2009.
\newblock Bpr: Bayesian personalized ranking from implicit feedback.
\newblock In {\em Proceedings of the twenty-fifth conference on uncertainty in
  artificial intelligence},  452--461.
\newblock AUAI Press.

\bibitem[\protect\citeauthoryear{Rendle, Freudenthaler, and
  Schmidt{-}Thieme}{2010}]{RendleFpmc}
Rendle, S.; Freudenthaler, C.; and Schmidt{-}Thieme, L.
\newblock 2010.
\newblock Factorizing personalized markov chains for next-basket
  recommendation.
\newblock In {\em {WWW} 2010, Raleigh, North Carolina, USA, April 26-30, 2010},
   811--820.

\bibitem[\protect\citeauthoryear{Ruxton}{2006}]{ruxton2006unequal}
Ruxton, G.~D.
\newblock 2006.
\newblock The unequal variance t-test is an underused alternative to student's
  t-test and the mann--whitney u test.
\newblock {\em Behavioral Ecology} 17(4):688--690.

\bibitem[\protect\citeauthoryear{Shani, Heckerman, and
  Brafman}{2005}]{shani2005mdp}
Shani, G.; Heckerman, D.; and Brafman, R.~I.
\newblock 2005.
\newblock An mdp-based recommender system.
\newblock {\em Journal of Machine Learning Research} 6(Sep):1265--1295.

\bibitem[\protect\citeauthoryear{Sun \bgroup et al\mbox.\egroup
  }{2019}]{sun2019bert4rec}
Sun, F.; Liu, J.; Wu, J.; Pei, C.; Lin, X.; Ou, W.; and Jiang, P.
\newblock 2019.
\newblock Bert4rec: Sequential recommendation with bidirectional encoder
  representations from transformer.
\newblock {\em arXiv preprint arXiv:1904.06690}.

\bibitem[\protect\citeauthoryear{Tang and Wang}{2018}]{Tangcaser}
Tang, J., and Wang, K.
\newblock 2018.
\newblock Personalized top-n sequential recommendation via convolutional
  sequence embedding.
\newblock In {\em Proceedings of the Eleventh ACM International Conference on
  Web Search and Data Mining}, WSDM '18,  565--573.
\newblock New York, NY, USA: ACM.

\bibitem[\protect\citeauthoryear{Tang \bgroup et al\mbox.\egroup
  }{2015}]{tang2015line}
Tang, J.; Qu, M.; Wang, M.; Zhang, M.; Yan, J.; and Mei, Q.
\newblock 2015.
\newblock Line: Large-scale information network embedding.
\newblock In {\em Proceedings of the 24th international conference on world
  wide web, WWW},  1067--1077.
\newblock International World Wide Web Conferences Steering Committee.

\bibitem[\protect\citeauthoryear{Wang \bgroup et al\mbox.\egroup
  }{2015}]{HRM2015}
Wang, P.; Guo, J.; Lan, Y.; Xu, J.; Wan, S.; and Cheng, X.
\newblock 2015.
\newblock Learning hierarchical representation model for nextbasket
  recommendation.
\newblock In {\em SIGIR},  403--412.
\newblock ACM.

\bibitem[\protect\citeauthoryear{Wang \bgroup et al\mbox.\egroup
  }{2019a}]{DBLP:conf/sigir/WangRMCMR19}
Wang, M.; Ren, P.; Mei, L.; Chen, Z.; Ma, J.; and de~Rijke, M.
\newblock 2019a.
\newblock A collaborative session-based recommendation approach with parallel
  memory modules.
\newblock In {\em {SIGIR} , Paris, France, July 21-25, 2019},  345--354.

\bibitem[\protect\citeauthoryear{Wang \bgroup et al\mbox.\egroup
  }{2019b}]{NGCF}
Wang, X.; He, X.; Wang, M.; Feng, F.; and Chua, T.-S.
\newblock 2019b.
\newblock Neural graph collaborative filtering.
\newblock In {\em SIGIR}.

\bibitem[\protect\citeauthoryear{Wu \bgroup et al\mbox.\egroup
  }{2019}]{shuwuSrgnn}
Wu, S.; Tang, Y.; Zhu, Y.; Wang, L.; Xie, X.; and Tan, T.
\newblock 2019.
\newblock Session-based recommendation with graph neural networks.
\newblock In {\em The Thirty-Third {AAAI} Conference on Artificial
  Intelligence},  346--353.

\bibitem[\protect\citeauthoryear{Xiao, Liang, and
  Meng}{2019}]{DBLP:conf/www/XiaoLM19}
Xiao, T.; Liang, S.; and Meng, Z.
\newblock 2019.
\newblock Hierarchical neural variational model for personalized sequential
  recommendation.
\newblock In {\em WWW 2019, San Francisco, CA, USA, May 13-17, 2019},
  3377--3383.

\bibitem[\protect\citeauthoryear{Xu \bgroup et al\mbox.\egroup
  }{2019}]{xu2019recurrent}
Xu, C.; Zhao, P.; Liu, Y.; Xu, J.; S~Sheng, V. S.~S.; Cui, Z.; Zhou, X.; and
  Xiong, H.
\newblock 2019.
\newblock Recurrent convolutional neural network for sequential recommendation.
\newblock In {\em WWW},  3398--3404.
\newblock ACM.

\bibitem[\protect\citeauthoryear{Ying \bgroup et al\mbox.\egroup
  }{2018}]{rexyingPinsage}
Ying, R.; He, R.; Chen, K.; Eksombatchai, P.; Hamilton, W.~L.; and Leskovec, J.
\newblock 2018.
\newblock Graph convolutional neural networks for web-scale recommender
  systems.
\newblock In {\em Proceedings of the 24th {ACM} {SIGKDD} International
  Conference on Knowledge Discovery {\&} Data Mining},  974--983.

\bibitem[\protect\citeauthoryear{Yuan \bgroup et al\mbox.\egroup
  }{2019}]{yuanfajiecnn}
Yuan, F.; Karatzoglou, A.; Arapakis, I.; Jose, J.~M.; and He, X.
\newblock 2019.
\newblock A simple convolutional generative network for next item
  recommendation.
\newblock In {\em Proceedings of the Twelfth {ACM} International Conference on
  Web Search and Data Mining, {WSDM} 2019, Melbourne, VIC, Australia, February
  11-15, 2019},  582--590.

\bibitem[\protect\citeauthoryear{Zhang \bgroup et al\mbox.\egroup
  }{2019}]{zhang2019next}
Zhang, S.; Tay, Y.; Yao, L.; Sun, A.; and An, J.
\newblock 2019.
\newblock Next item recommendation with self-attentive metric learning.
\newblock In {\em Thirty-Third AAAI Conference on Artificial Intelligence},
  volume~9.

\bibitem[\protect\citeauthoryear{Zhou \bgroup et al\mbox.\egroup
  }{2017}]{zhou2017scalable}
Zhou, C.; Liu, Y.; Liu, X.; Liu, Z.; and Gao, J.
\newblock 2017.
\newblock Scalable graph embedding for asymmetric proximity.
\newblock In {\em Thirty-First AAAI Conference on Artificial Intelligence,
  AAAI}.

\end{thebibliography}

\end{document}